\documentclass{elsart}
\usepackage{amsmath,amssymb}
\usepackage{cite}
\usepackage{graphicx}
\allowdisplaybreaks

\newcommand{\sla}[1]{/\!\!\!#1}

\def\lsim{\raise0.3ex\hbox{$\;<$\kern-0.75em\raise-1.1ex\hbox{$\sim\;$}}}
\def\gsim{\raise0.3ex\hbox{$\;>$\kern-0.75em\raise-1.1ex\hbox{$\sim\;$}}}

\begin{document}

\hfill{YITP-SB-10--23}

\begin{frontmatter}
\title{Scrutinizing the $ZW^+W^-$ vertex at the Large Hadron Collider at 7 TeV}
\author{O.\ J.\ P.\ \'Eboli}
\address{Instituto de F\'{\i}sica,
             Universidade de S\~ao Paulo, S\~ao Paulo -- SP, Brazil.}

\author{J.\ Gonzalez-Fraile}
\address{Departament d'Estructura i Constituents de la Mat\`eria and
  ICC-UB, Universitat de Barcelona, 647 Diagonal, E-08028 Barcelona,
  Spain}

\author{M.\ C.\ Gonzalez-Garcia}
\address{Instituci\'o Catalana de Recerca i Estudis Avan\c{c}ats
  (ICREA), Departament d'Estructura i Constituents de la Mat\`eria and
  ICC-UB, Universitat de Barcelona, 647 Diagonal, E-08028 Barcelona,
  Spain}
\address{C.N.~Yang Institute for Theoretical Physics,
  SUNY at Stony Brook, Stony Brook, NY 11794-3840, USA}

\begin{abstract}
  We analyze the potential of the CERN Large Hadron Collider running
  at 7 TeV to search for deviations from the Standard Model
  predictions for the triple gauge boson coupling $ZW^+W^-$ assuming
  an integrated luminosity of 1 fb$^{-1}$.  We show that the study of
  $W^+W^-$ and $W^\pm Z$ productions, followed by the leptonic decay
  of the weak gauge bosons can improve the present sensitivity on the
  anomalous couplings $\Delta g_1^Z$, $\Delta \kappa_Z$, $\lambda_Z$,
  $g_4^Z$, and $\tilde{\lambda}_Z$ at the $2\sigma$ level.
\end{abstract}
\end{frontmatter}


Recently the CERN Large Hadron Collider (LHC) started a run with
center-of-mass energy of 7 TeV and plans to accumulate an integrated
luminosity of $\simeq 1$ fb$^{-1}$. This new high energy frontier
allow us to further test the Standard Model (SM), as well as, to check
for its possible extensions. In particular, within the framework of
the SM, the structure of the trilinear and quartic vector--boson
couplings is completely determined by the $SU(2)_L \times U(1)_Y$
gauge symmetry. Thus the study of these interactions can either lead
to an additional confirmation of the model or give some hint on the
existence of new phenomena at a higher scale~\cite{anomalous}.  The
triple gauge--boson vertices (TGV's) have been probed directly at the
Tevatron~\cite{tevatron} and LEP~\cite{lep} through the production of
vector--boson pairs and the experimental results agree with the SM
predictions, see Table \ref{tab:bounds}. Moreover, TGV's contribute at
the one--loop level to the $Z$ physics and consequently they can also
be indirectly constrained by precision electroweak
data~\cite{indirect}.  At the LHC, the TGV's will be subject to a more
severe scrutiny via the production of electroweak gauge boson pairs,
{\em e.g.} $W \gamma$ and $WZ$. Running at 14 TeV center-of-mass
energy and with 30--100 $fb^{-1}$ integrated luminosity it will probe
these couplings at the few percentage level; see
Ref.\cite{Dobbs:2005ev} for a recent update.

In this work we assess the potential of the LHC already running at 7
TeV to probe deviations from the SM prediction for the $ZW^+W^-$
interaction through the reactions
\begin{eqnarray}
p p &&\to W^+ W^- \to \ell^+ \ell^{\prime -} \sla{E}_T 
\label{ppww}
\\
p p && \to W^\pm Z \to \ell^{\prime\pm} \ell^{+}\ell^{-} \sla{E}_T
\label{ppwz}
\end{eqnarray}
where $\ell^{(\prime)} = e$ or $\mu$. 

The most general form of the $ZW^+W^-$ vertex compatible with Lorentz
invariance is given by the effective Lagrangian \cite{Hagiwara:1986vm}
\begin{eqnarray} 
{\mathcal L}_{\text{eff}} /g_{WWZ} = && + i g_1^Z \left (
    W^\dagger_{\mu\nu} W^\mu Z^\nu - W^\dagger_\mu W^{\mu\nu} Z_\nu \right ) 
+ i \kappa_Z W^\dagger_\mu W_\nu Z^{\mu\nu}
\nonumber
\\
&& 
+ i \frac{\lambda_Z}{M^2_W} W^\dagger_{\rho \mu} W^\mu_\nu Z^{\nu \rho}
 + g_5^Z \epsilon^{\mu\nu\rho\sigma} (W^\dagger_\mu \partial_\rho W_\nu -
\partial_\rho W^\dagger_\mu W_\nu ) Z_\sigma
\label{leff}
\\
&&
- g_4^Z W^\dagger_\mu W_\nu (\partial^\mu Z^\nu + \partial^\nu Z^\mu)
 + i \tilde{\kappa}_Z W^\dagger_\mu W_\nu \tilde{Z}^{\mu\nu}
+ i \frac{\tilde{\lambda}_Z}{M_W^2} W^\dagger_{\sigma\mu} W^\mu_\nu 
\tilde{Z}^{\nu\sigma}
\nonumber
\end{eqnarray}
where $Z^{\mu\nu} = \partial^\mu Z^\nu -\partial^\nu Z^\mu$ and
$\tilde{Z}^{\mu\nu} = \frac{1}{2} \epsilon^{\mu\nu\rho\sigma}Z_{\rho\sigma}$.
$g_{WWZ} = -e \cot \theta_W$ and $\theta_W$ is the weak mixing
angle. The couplings $g_1^Z$, $\kappa_Z$ and $\lambda_Z$ are C and P
conserving, $\tilde{\kappa}_Z$ and $\tilde{\lambda}^Z$ are P odd and
violate CP, while $g_4^Z$ violates C and CP and $g_5^Z$ violates C and
P but is CP conserving. In the SM $g_1^Z = \kappa_Z =1$ and $\lambda_Z
= g_4^Z =g_5^Z = \tilde{\kappa}_Z = \tilde{\lambda}_Z = 0$.

\begin{table}
\begin{center}
 \begin{tabular}{||l|r|c|r|r||}\hline
    couplings       & PDG bounds             & indirect limits
       & Unit. $W^+ W^-$   &  Unit. $W^\pm Z$  
\\ \hline                                                                   
    $\Delta g_1^Z$    & $-0.016^{+0.022}_{-0.019}$ & $[-0.051~,~0.0092]$
   &     2.7     &  0.22       
\\  \hline                                                                  
    $\Delta \kappa_Z$      & $-0.076^{+0.059}_{-0.056}$ & $[-0.050~,~0.0039]$
   &    0.22     &   3.5       
\\ \hline                                                                   
    $\lambda_Z$     & $-0.088^{+0.060}_{-0.057}$ & $[-0.061~,~0.10]$ 
  &    0.15     &  0.14       
\\ \hline                                                                   
    $g_5^Z$         & $-0.07 \pm 0.09$         & $[-0.085~,~0.049]$ 
     &     2.7     &   1.7      
\\ \hline                                                                   
    $g_4^Z$         & $-0.30\pm0.17$          & ---
&     2.7     &  0.22      
\\ \hline                                                                   
$\tilde{\kappa}_Z$  & $-0.12^{+0.06}_{-0.04}$   & ---
    &   2.7     &   3.5        
\\ \hline                                                                   
$\tilde{\lambda}_Z$ & $-0.09\pm0.07$          & ---
   &  0.15     &  0.14        
\\ \hline
 \end{tabular}
\end{center}
\vglue 0.3cm
\caption{Available limits on the anomalous TGV  couplings.
  The first column contains a compilation of the direct searches
  performed by the Particle Data Group~\cite{Amsler:2008zzb}.
  The indirect bounds are presented in the second column~\cite{indirect} where
  the entries not evaluated in the literature are marked as ---.
  The third and fourth columns contain the  bounds
  derived from the processes $q q \to W^+W^-$ and $W^\pm Z$ ~\cite{unitarity}  
  imposing that unitarity is satisfied for energies below 2 TeV.}
\label{tab:bounds}
\end{table}



In presence of these anomalous couplings the cross sections for the
processes  $p p \to \ell^+ \ell^{\prime -} \sla{E}_T$ and 
$p p \to \ell^\pm \ell^{\prime +}\ell^{\prime -} \sla{E}_T$
take the form
\begin{equation}
\sigma=\sigma_{\text{SM}}+ \sum_{i} \sigma_{\text{int}}^i ~g^i_{\text{ano}} 
+\sum_{i, j \geq i}\sigma_{\text{ano}}^{ij} ~g_{\text{ano}}^i
~g_{\text{ano}}^j  \;\; , 
\label{def:sigma}
\end{equation}
where $\sigma_{\text{SM}}$, $\sigma_{\text{int}}^i$, and
$\sigma_{\text{ano}}^{ij}$ are, respectively, the SM contribution, the
interference between the SM and the anomalous contribution, and the
pure anomalous ones.  For the CP violating couplings
$\sigma_{\text{int}}^i$ vanishes.

SM contributions to $p p \to \ell^+ \ell^{\prime -} \sla{E}_T$ include 
electroweak (EW) processes leading to this final state -- such us 
$W^+W^-$ production or $ZZ$ production with one $Z$ decaying in charged leptons and the other in
neutrinos --  and $t\bar{t}$ production with the top quarks decaying
semi-leptonicaly.  For $p p \to \ell^\pm \ell^{\prime +}\ell^{\prime
  -} \sla{E}_T$ the main SM backgrounds are the EW production of
$W^\pm Z$ pairs and $ZZ$ production with the subsequent decays of the
$Z$'s into leptons when one charged lepton escapes detection. An
additional background comes from $t \bar{t}$ production if the
semi-leptonic decay of a $b$ gives rise to an isolated charged lepton.

The signal and backgrounds were simulated at the parton level with
full tree level matrix elements generated with the package
MadEvent~\cite{madevent} conveniently modified to include the
anomalous TGV's.  We employed CTEQ6L parton distribution functions
\cite{cteq6l} throughout. We took the electroweak parameters to be
$\alpha_{em} = 1/132.51$, $m_Z = 91.188$ GeV, $m_W = 80.419$ GeV, and
$\sin^2 \theta_W = 0.222$, which was obtained imposing the tree level
relation $\cos \theta_W = m_W/m_Z$.  We simulated experimental
resolutions by smearing the energies (but not directions) of all final
state charged leptons with a Gaussian error $\Delta(E)/E =
0.02/\sqrt{E}$. We also included in our analysis a 90\% lepton
detection efficiency.

We began our analysis of processes (\ref{ppww}) and (\ref{ppwz}) by
imposing some basic acceptance cuts for the charged leptons and 
missing energy
\begin{equation}
   p_T^\ell \ge  10 \hbox{ GeV} \;\;\;,\;\;\;
   |\eta_\ell| < 2.5 \;\;\;,\;\;\;
  \Delta R_{\ell\ell} \ge 0.4  \;\;\;,\;\;\; \sla{p}_T\geq 10\; {\rm GeV}
\label{basiccuts}
\end{equation}
where $\eta_\ell$ is the charged lepton pseudo-rapidity. 

For $p p \to \ell^+ \ell^- \sla{E}_T$ events with the two leptons of
the same flavor we further required that the lepton pair invariant
mass ($M_{\ell\ell}$) is not compatible with a $Z$ production, {\em
  i.e.}
\begin{equation}
  | M_{\ell\ell} - M_Z | > 10 \hbox{ GeV} \; .
\label{zveto}
\end{equation}
Furthermore the top quark pairs are a potentially large background due
to its production by strong interactions. To further suppress these
events we vetoed the presence of central jets with
\begin{equation}
   p_T^j  > 20 \hbox{ GeV} \;\;\;\;\hbox{ and }\;\;\;\;
   | \eta_j| < 3 \; .
\label{jetveto} 
\end{equation}

For $p p \to \ell^{\prime \pm} \ell^{+}\ell^{-} \sla{E}_T$ 
in the case with only a pair of same flavor different sign 
leptons (this is, $\ell^\prime  \neq \ell$) we demanded that the invariant
mass of the equal flavor lepton pair is compatible with the 
$Z$ mass, {\em i.e.}
\begin{equation}
   |M_{\ell \ell} - M_Z| < 10 \hbox{ GeV} \;, 
\label{isaz}
\end{equation}

The presence of just one neutrino in the final state of this channel
permits the reconstruction of its momentum by imposing the transverse
momentum conservation and requiring that the invariant mass of the
third lepton and the neutrino is the $W$ mass
\begin{equation}
  M_{\ell^\prime \nu} = M_W \;.
\label{isaw}
\end{equation}
This procedure exhibits a twofold ambiguity on the neutrino
longitudinal momentum. In our analysis we kept only events that
possess a solution to the neutrino momentum.


Conversely when the three leptons have the same flavor we demanded
that one opposite sign lepton pair satisfies (\ref{isaz}) and the
third lepton and the missing transverse momentum reconstructs a $W$ as
in (\ref{isaw}). We further required that the invariant mass of the
third lepton and the lepton of opposite charge used to reconstruct the
$Z$ is not compatible with a $Z$, therefore complying with (\ref{zveto}).  The
top pair background to
$p p \to \ell^{\prime \pm} \ell^{+}\ell^{-} \sla{E}_T$ 
after cuts \eqref{basiccuts} and (\ref{isaz})--(\ref{isaw}) 
(plus (\ref{zveto}) for $\ell'=\ell$) is
already very suppressed since it requires that one of the isolated
leptons originates from a $b$ quark semi-leptonic decay.  Vetoing any
central jet activity, as in Eq.~(\ref{jetveto}) renders the $t
\bar{t}$ cross section negligible.

\begin{table}[htb]
\begin{center}
 \begin{tabular}{|c|c||c|c|c|c||c|c|c|}
\hline
 \multicolumn{2}{|c||}
{$\sigma_{\text{SM}}$ (fb)}
& \multicolumn{4}{c||}
{$\begin{array}{l}\sigma_{\text{ano}}\;{\rm  (fb)}
\\ \sigma_{\text{int}}\;{\rm  (fb})\end{array}$
}
&\multicolumn{3}{c|}
{$\begin{array}{l}\sigma_{\text{ano}}\;{\rm (fb)}\\ 
\Delta\sigma_{\text{ano}}\;{\rm  (fb)}\end{array}$}
\\
\hline
\multicolumn{9}{|c|} {$pp \to \ell^+\ell^{\prime -} \sla{E}_T$ }
\\
\hline $l^+\nu_ll'^-\nu_{l'}$ 
& $t \bar{t}$
& $\Delta g_1^Z$ 
& $\Delta \kappa_Z$ 
&  $\lambda_Z$ 
& $g_5^Z$ 
& $g_4^Z$ 
& $\tilde{\kappa}_Z$ 
& $\tilde{\lambda}_Z$ 
\\
\hline
824.
& 11.1
& 254.
& 2540.
& 5750.
& 163.
& 219.
& 412.
& 6030.
\\ \hline
&
&  -55.7
&  -166.
&  -22.1   
&   15.1
&  68.8
&  -89.2
&   152. 
\\ \hline\hline
\multicolumn{9}{|c|}
{$p p \to \ell^{\prime \pm} \ell^{+}\ell^{-} \sla{E}_T$}
\\
\hline 
$\ell^+\ell^-\ell^{\prime \pm} \nu$
& $ZZ$ 
& $\Delta g_1^Z$ 
& $\Delta \kappa_Z$ 
&  $\lambda_Z$ 
& $g_5^Z$ 
& $g_4^Z$ 
& $\tilde{\kappa}_Z$ 
& $\tilde{\lambda}_Z$ 
\\ \hline
63.0
& 2.32
&   1280.
&   65.4
&  2290.
&    391.
&  1020.
& 77.6
& 2390.
\\\hline
& 
&  -106.
&  -21.3
&  -24.3
&   -7.2
&  -20.2
&  -2.2
&  -10.0
\\\hline
\end{tabular}
\end{center}
\vglue 0.2cm
\caption{
  Cross sections for the process   $pp \to \ell^+\ell^{\prime -} \sla{E}_T$
  after the cuts (\ref{basiccuts})--(\ref{jetveto}) and 
$p p \to \ell^{\prime \pm} \ell^{+}\ell^{-} \sla{E}_T$ 
  after the cuts  (\ref{basiccuts}) and (\ref{jetveto})--(\ref{isaw}) 
 (plus \eqref{zveto} for $\ell'=\ell$). 
  We denote by    $ZZ$ the process $pp\to \ell^+\ell^-\ell^{\prime \pm} 
  [\ell^{\prime \mp}]$   where  $[\ell^\pm]$ is a charged lepton that escapes 
  detection.   In all cases the results include the charge lepton detection 
  efficiency. For the CP violating couplings we provide the result for
  $\Delta \sigma_{\text{ano}}$, see Eq.~(\ref{deltasigCP}).}
\label{tab:resww}
\end{table}

We present in Table~\ref{tab:resww} the cross sections of the SM
backgrounds and anomalous contributions to process $pp \to
\ell^+\ell^{\prime -} \sla{E}_T$ after the cuts
(\ref{basiccuts})---(\ref{jetveto}) and  
$p p \to \ell^{\prime \pm} \ell^{+}\ell^{-} \sla{E}_T$ 
after the cuts (\ref{basiccuts}) and
(\ref{jetveto})--(\ref{isaw}) (plus (\ref{zveto})
for $\ell'=\ell$).  For $pp \to \ell^+\ell^{\prime -} \sla{E}_T$ the
cut (\ref{jetveto}) is very important to tame the dangerous $t\bar{t}$
background whose cross section is 3.9 pb when we remove this cut.  For
simplicity we have only considered one non-vanishing anomalous vertex
at a time.  This simplifying hypothesis can be consistently made when
the integration of the heavy degrees of freedom associated to a new
physics leads to scenario where the $SU(2)_L \times U(1)_Y$ symmetry
is realized non-linearly in the low energy effective Lagrangian. If
the low energy remains of the new physics are described by a linear
realization of the $SU(2)_L \times U(1)_Y$ there will be relations
between the anomalous TGV's; see, for instance,
reference~\cite{linear}.

\begin{figure}
\centering
\includegraphics[width=0.9\textwidth]{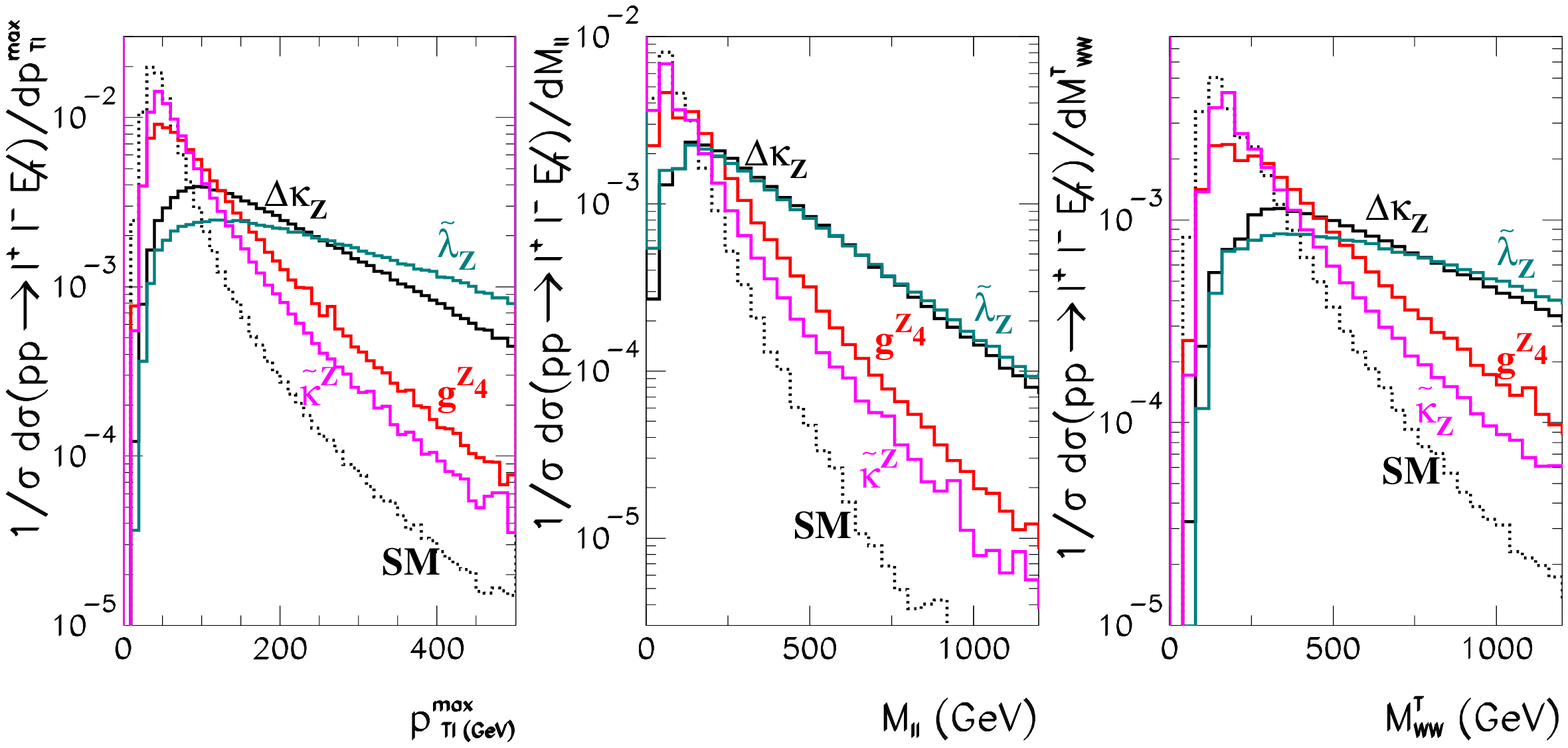}
\includegraphics[width=0.9\textwidth]{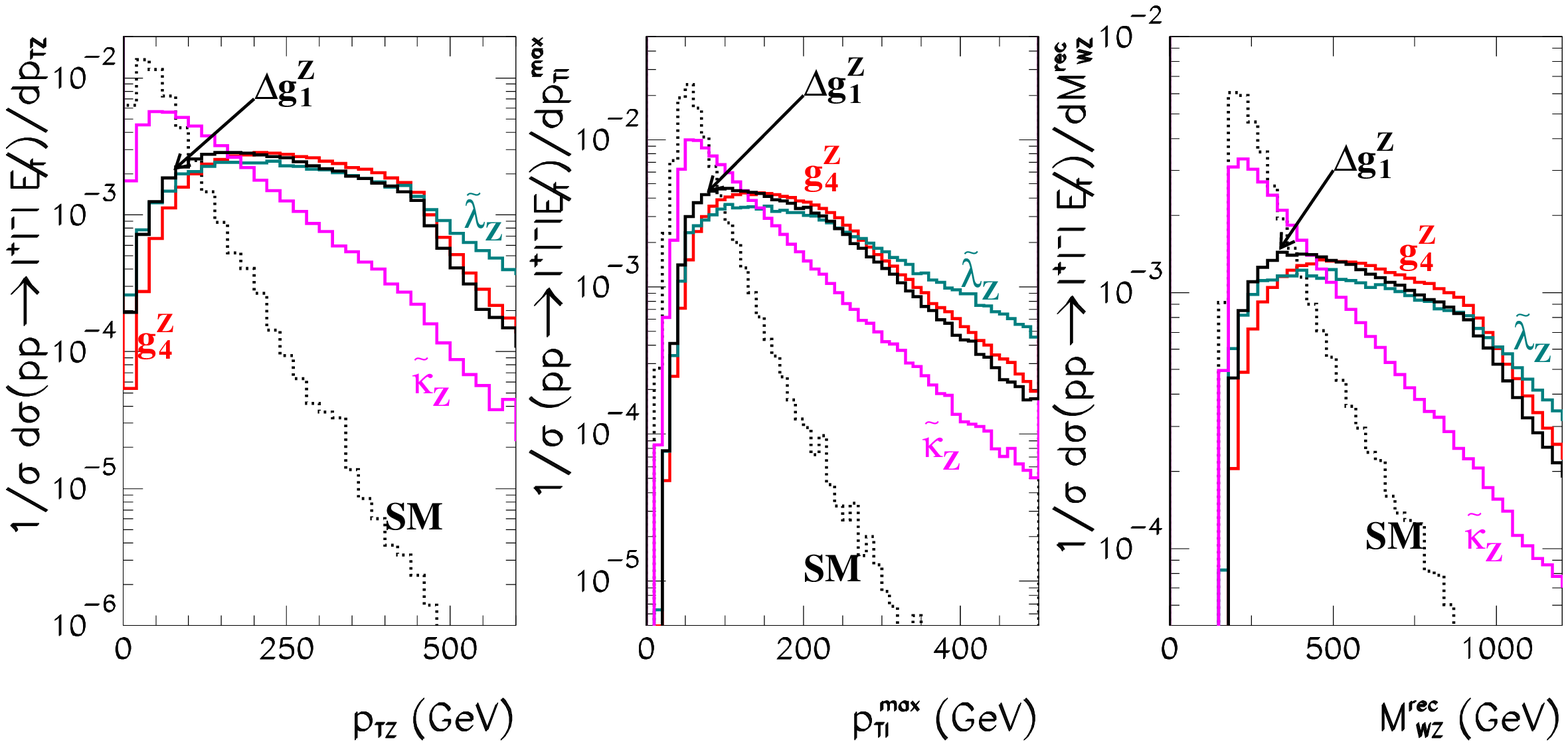}
\caption{Normalized spectra for some relevant kinematic variables for
  the SM and some of the anomalous TGV's for $pp \to
  \ell^+\ell^{\prime -} \sla{E}_T$ (upper panels) and 
$p p \to \ell^{\prime \pm} \ell^{+}\ell^{-} \sla{E}_T$ 
(lower panels).
  The upper panels show the distribution in transverse momentum of the
  hardest lepton (left panel), dilepton invariant mass (central panel)
  and reconstructed $WW$ transverse invariant mass (right panel). The
  lower panels show the transverse momentum of the $Z$ (left panel),
  hardest lepton transverse momentum (central panel), and
  reconstructed $WZ$ transverse invariant mass (right panel).}
  \label{fig:distri}
\end{figure}

The normalized spectra of the SM and the anomalous contributions for
some relevant kinematic variables are displayed in Figure
\ref{fig:distri}.  For $pp \to \ell^+\ell^{\prime -} \sla{E}_T$ we
have defined the transverse mass $M_T^{WW}$ as:
\begin{equation}
  M_T^{WW} 
= \biggl[ \left( \sqrt{(p_T^{\ell^+\ell^{\prime -}})^2 + m^2_{\ell^+ \ell^{\prime -}}} 
+ \sqrt{\sla{p_T}^2 + m^2_{\ell^+\ell^{\prime -} }} \right)^2 \biggr .
\biggl . - (\vec{p}_T^{~\ell^+\ell^{\prime -}} + \vec{\sla{p_T}}  )^2 \biggr]^{1/2} 
\label{mtww}
\end{equation}
where $\vec{\sla{p_T}}$ is the missing transverse momentum vector,
$\vec{p}_T^{~\ell^+\ell^{\prime -}}$ is the transverse momentum of the
pair $\ell^+ \ell^{\prime -}$ and $m_{\ell^+\ell^{\prime -}}$ is the
$\ell^+ \ell^{\prime -}$ invariant mass. 

For $p p \to \ell^{\prime \pm} \ell^{+}\ell^{-} \sla{E}_T$ we
define $p_{TZ}$ as the transverse momentum of the opposite sign equal
flavor leptons verifying (\ref{isaz}). Furthermore, it is possible to
reconstruct the neutrino momentum, and consequently, we can evaluate
the total $\ell\ell\ell\nu$ invariant mass (which we label $M_{WZ}$)
that takes two possible values for each event.  In the lower right
panel of Fig.~\ref{fig:distri} we show the distribution in this
variable where each of the solutions have been given weight $1/2$.

Figure~\ref{fig:distri} illustrates the well-known fact that the
anomalous contributions enhance the cross section at higher collision
energies (eventually leading to perturbative unitarity violation) and
that this behaviour can be well traced by either $p_{T \ell}^{max}$,
$M_{\ell\ell}$, $M^T_{WW}$, $p_{TZ}$ or $M_{WZ}$ respectively.

In order to extract the attainable sensitivity on anomalous TGV we
analyzed for each kinematic variable shown in Fig.~\ref{fig:distri}
the choice of cut that maximizes the sensitivity for deviations in the
TGV's.  Given the limited statistics of the 7 TeV LHC run we do not
attempt to make a fit to the distributions and use instead as unique
variable the total number of observed events above a certain minimum
cut for each of the variables. In each case we assumed that the total
number of observed events is the one predicted by the SM at integrated
luminosity of 1 fb$^{-1}$. The corresponding statistical uncertainty
were obtained using Poisson or Gaussian statistics depending on
whether the expected number of SM events was smaller or larger than
20.  We performed our analysis of the channels $pp \to
\ell^+\ell^{\prime -} \sla{E}_T$ and 
$p p \to \ell^{\prime \pm} \ell^{+}\ell^{-} \sla{E}_T$ 
 independently.

\begin{figure}
  \centering
  \includegraphics[width=0.85\textwidth]{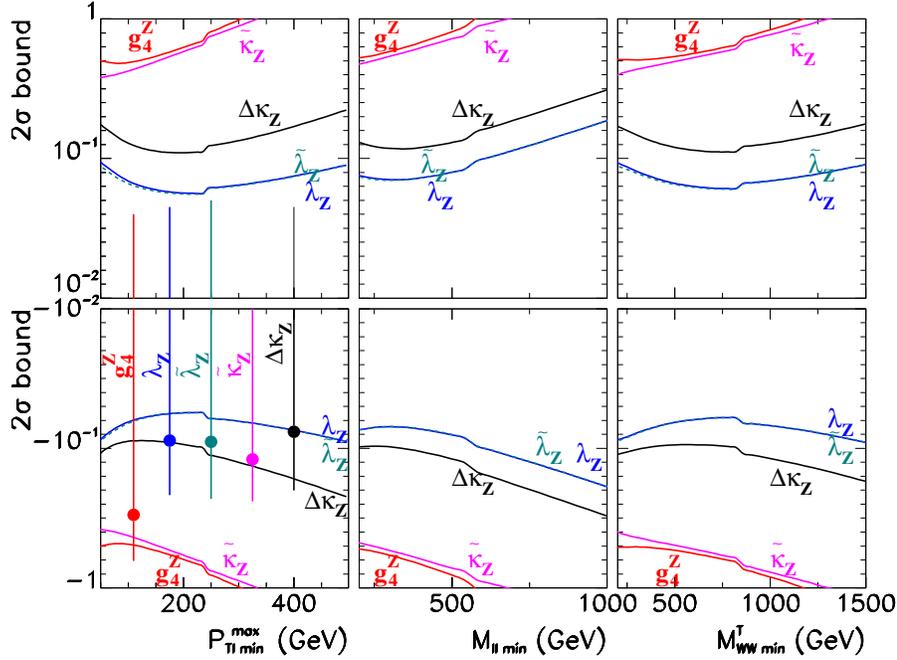}
  \caption{Dependence of the upper (top panels) and lower (lower
    panels) $2\sigma$ bounds achievable from the study of $pp \to
    \ell^+\ell^{\prime -} \sla{E}_T$ as a function of the cut on the
    minimum value on the hardest lepton transverse momentum (left
    panels), the dilepton invariant mass (central panels), and the
    reconstructed $WW$ transverse invariant mass (right panels). The
    dashed curves correspond to $\tilde\lambda_Z$ and are almost
    indistinguishable from the full blue lines corresponding to
    $\lambda_Z$ . The presently allowed $2\sigma$ ranges are indicated
    by the vertical lines in the left panels.}
  \label{fig:boundww}
\end{figure}

We depict in Figure \ref{fig:boundww} the achievable $2\sigma$ limits
from $pp \to \ell^+\ell^{\prime -} \sla{E}_T$ on some of the anomalous
TGV as a function of the the minimum cut on the maximum transverse
momentum of the leptons (left panels), dilepton invariant mass
(central panels) and minimum reconstructed transverse invariant mass
($M_{WW}^T$) in the right panels.  As we can see from this figure the
couplings $\Delta \kappa_Z$, $\tilde{\lambda}_Z$ and $\lambda_Z$
possess a mild dependence on the kinematic cut while $g_4^Z$ and
$\tilde{\kappa}_Z$ experience larger changes with the variation of the
cuts.  This can be understood from Fig.~\ref{fig:distri} that shows
that $g_4^Z$ and $\tilde{\kappa}_Z$ distributions decrease much faster
that the ones for $\tilde{\lambda}_Z$. We find that maximum
sensitivity for any of the anomalous TGV is obtained from a minimum
cut in transverse momentum of the hardest lepton with the optimum cut
ranging between 50--200 GeV depending on the anomalous coupling
considered. The corresponding attainable 2$\sigma$ bounds are listed
in Table~\ref{tab:wwlimits}. Our results show that this channel can
tighten the present direct bounds on $\Delta \kappa_Z$, $\lambda_Z$,
$g_4^Z$, and $\tilde{\lambda}_Z$.

\begin{table}
\begin{center}
  \begin{tabular}{|c||c|c|c|c|}  \hline
          & \multicolumn{2}{c|}{$W^+W^-$  $2\sigma$ limits}  
& \multicolumn{2}{c|}{$W^\pm Z$  $2\sigma$ limits }
\\
\hline 
& No form factor & $\Lambda=3$ TeV  
& No form factor & $\Lambda=3$ TeV \\  \hline
    $\Delta g_1^Z$ 
& $[-0.33\;,\; 0.56]$& $[-0.35\;,\; 0.59]$ 
& $[-0.055 \;,\; 0.094]$ & $[-0.061 \;,\; 0.11]$
\\
\hline
$\Delta \kappa_Z$ 
& $[-0.088 \;,\;0.11]$ & $[-0.10 \;,\;0.14]$ 
& $[-0.27 \;,\;  0.55]$ & $[-0.29 \;,\;  0.61]$
\\
\hline
    $\lambda_Z$   
& $[-0.055 \;,\;  0.056]$&  $[-0.074 \;,\;  0.075]$
& $[-0.051\;,\;  0.054]$ & $[-0.060\;,\;  0.064]$
\\
\hline
    $g_5^Z$    
& $[-0.53 \;,\; 0.51]$& $[-0.56 \;,\; 0.55]$
& $[-0.18 \;,\; 0.19]$ & $[-0.19 \;,\; 0.20]$
\\
\hline
    $g_4^Z$    
& $[-0.48 \;,\;  0.48]$  &  $[-0.51 \;,\;  0.51]$    
&$[-0.080 \;,\; 0.080]$ & $[-0.091 \;,\; 0.091]$
\\
\hline
    $\tilde{\kappa}_Z$ 
& $[-0.38 \;,\;0.38]$& $[-0.39 \;,\;0.39]$
& $[-0.40 \;,\; 0.40]$& $[-0.42 \;,\; 0.42]$
\\
\hline
   $\tilde{\lambda}_Z$ 
& $[-0.055 \;,\;  0.055]$&  $[-0.074 \;,\;  0.074]$  
& $[-0.053 \;,\; 0.053]$ & $[-0.062 \;,\; 0.062]$
\\
\hline
\end{tabular}
\end{center}
\vglue 0.2cm
\caption{Attainable $2\sigma$ bounds on anomalous TGV at the LHC at 7 TeV 
  with 1 fb$^{-1}$.} 
\label{tab:wwlimits}
\end{table}

\begin{figure}
  \centering
  \includegraphics[width=0.85\textwidth]{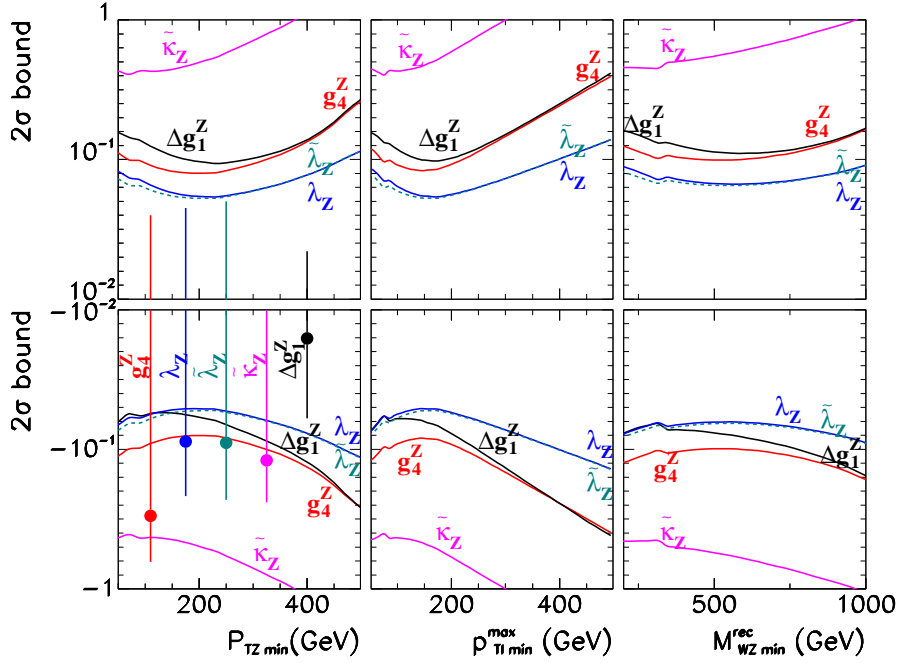}
  \caption{ Dependence of the upper (top panels) and lower (lower
    panels) $2\sigma$ bounds from 
$p p \to \ell^{\prime \pm} \ell^{+}\ell^{-} \sla{E}_T$ 
 a function of the cut on the minimum value  on the 
$Z$ transverse 
   momentum (left panels),  the  hardest
    lepton transverse momentum (central panels),
 and the reconstructed $WZ$  
    invariant mass (right panels). 
The dashed curves correspond to 
     $\tilde\lambda_Z$ and are almost indistinguishable from the full blue
lines corresponding to $\lambda_Z$ 
. The presently allowed $2\sigma$
    ranges are indicated by the vertical lines in the left panels.}
  \label{fig:boundwz}
\end{figure}

Correspondingly we show in Figure~\ref{fig:boundwz} the bounds from
$p p \to \ell^{\prime \pm} \ell^{+}\ell^{-} \sla{E}_T$ 
as a function
of the minimum cut in either the $Z$ transverse momentum (left
panels), the hardest lepton transverse momentum (central panels), and
the reconstructed $WZ$ invariant mass (right panels).  We find that a
minimum cut in either of the transverse momentum variables (that of
the $Z$ or the hardest lepton $p_T$) leads to the best
sensitivity. The corresponding attainable 2$\sigma$ bounds are also
listed in Table~\ref{tab:wwlimits} that shows that this channel can
improve the present direct constraints on the couplings $\Delta
g_1^Z$, $\lambda_Z$, $g_4^Z$ and $\tilde{\lambda}_Z$.

So far we have applied the same type of analysis to CP conserving or
CP violating couplings. For these last ones their CP breaking nature 
can be addressed constructing  some  CP-odd  or $\hat T$-odd 
observable by weighting the events with the sign of the relevant cross 
product of the measured momenta.  For example, following 
Refs.\cite{Han,Kum} we can define  
\begin{eqnarray}
\Xi_\pm &\equiv&
\mbox{sign}
\left[\left(\vec{p}_{\ell^+}-\vec{p}_{\ell^{\prime -}}\right)^z\right]\,
\mbox{sign}\left(\vec{p}_{\ell^+} \times\vec{p}_{\ell^{\prime -}}\right)^z
\;\;\; \;\;\; \;\; \,
\hfill
{\rm for}\; 
p p \to  \ell^{+}\ell^{\prime -} \sla{E}_T
\, ,  \label{ha} \\
\Xi_\pm &\equiv&\mbox{sign}(p^z_Z)\, \mbox{sign}(p_{\ell^\prime}\times p_Z)^z
\;\;\; \;\;\; \;\;\; \;\;\; \;\;\; 
\;\;\; \;\;\; \;\;\; \;\;\; \;\; 
\hfill {\rm for }\; p p \to \ell^{\prime \pm} \ell^{+}\ell^{-} \sla{E}_T
\, ,
\label{ku}
\end{eqnarray}
where $z$ is the collision axis. The CP-violating couplings  give a 
non-vanishing contribution to the sign-weighted cross section 
\begin{equation}
 g^i_{\text{ano}} \, \Delta \sigma^{i}_{\text{ano}} 
\equiv \int d\sigma  \,\Xi_\pm  \; . 
\label{deltasigCP}
\end{equation} 
We present in Table \ref{tab:resww} the values of the corresponding
sign-weighted cross sections. The resulting number of sign-weighted
events has to be compared with the statistical fluctuations from the
SM events (which are sign symmetric).  We find that given the existing
bounds on $\tilde{\kappa}_Z$, $\tilde{\lambda}^Z$ and $g_4^Z$, the
study of these events at the 7 TeV run of LHC is not precise enough to
provide information on the CP properties of the anomalous couplings.

It is well known that the introduction of anomalous couplings spoils
delicate cancellations in scattering amplitudes, leading to their
growth with energy and, eventually, to unitarity violation above a
certain scale $\Lambda$. The way to cure this problem that is being
used in the literature is to introduce an energy dependent form factor
that dumps the anomalous scattering amplitude growth at high energy,
such as
\begin{equation}
    \frac{1}{(1 + \frac{\hat{s}}{\Lambda^2})^2}
\label{eq:formfactor}
\end{equation}
where $\sqrt{\hat{s}}$ is the center--of--mass energy
of the $WW$ or $WZ$ pair.   
%
Here we advocate that the need to introduce a form factor at the 7 TeV
run of LHC is marginal because the center-of-mass energy for the
contributing sub-process in (\ref{ppww}) and (\ref{ppwz}) is $\lesssim
2$ TeV, and the unitarity bounds on the anomalous TGV steaming from
these processes are much weaker than the ones that we obtain; see the
fourth and fifth columns of Table~\ref{tab:bounds}.  In principle one
may worry about the corresponding unitarity violation in longitudinal
$VV$ ($V=W^\pm$ or $Z$) scattering which can lead to stronger bounds
on the TGV since they can lead to a scattering amplitude which grows
as $\hat s^2$.  However, the actual energy behaviour of the scattering
amplitude in longitudinal gauge boson scattering depends strongly on
the assumptions about the quartic gauge boson
couplings~\cite{unitWW}. In particular, if there is a mechanism
relating the quartic and triple anomalous contributions the $VV$
scattering unitarity bounds turn out to be similar to the ones in
reference~\cite{unitarity}. Altogether we find that within the bounds
that we derive, unitarity is held up to $\sqrt{\hat s} \simeq 3$ TeV.
As a final consistency check we derive the bounds obtained if a form
factor (\ref{eq:formfactor}) was included with $\Lambda=3$ TeV. We
show in Table~\ref{tab:wwlimits} the changes in the $2\sigma$
sensitivity.

Our analysis leaves some room for improvement. For instance,
we considered only one kinematic distribution to extract the bounds,
leaving out the possibility of optimizing the analysis for joint
distributions or a binned maximum likelihood fit. Moreover, our
calculations were carried out at the parton level with the lowest
order perturbation theory. Certainly a full Monte Carlo analysis
taking into account detector simulation, as well as, NLO
QCD~\cite{qcdnlo} and EW~\cite{ewnlo} is in order. Although QCD NLO
corrections are potentially dangerous due to changes in $p_T$
distributions, our jet veto cut (\ref{jetveto}) is enough to guarantee
that the attainable limits are not significantly altered
~\cite{qcdnlo,anonlo}.  In brief, we anticipate that our results should give
a fair estimate of the LHC potential to study the $ZW^+W^-$ vertex.

Summarizing,  we have shown that the study of the processes (\ref{ppww}) and
(\ref{ppwz}) at the LHC with a center--of--mass energy of 7 TeV 
and an integrated luminosity of 1 fb$^{-1}$ can
improve the presently available direct limits on the $Z$ anomalous
couplings $\Delta g_1^Z$, $\Delta \kappa_Z$, $\lambda_Z$, $g_4^Z$, and
$\tilde{\lambda}_Z$. 
Due to the small integrated luminosity predicted
for this initial run, the limits on these couplings will be only slightly
more stringent than the present available ones. Nevertheless, the more
precise results on $\Delta g_1^Z$ and $\lambda_Z$ will start to
compete with the indirect limits coming from precision measurements;
see Table~\ref{tab:bounds}.


\vskip 0.3cm 
This work is supported by USA-NSF grant PHY-0653342, by
Spanish grants from MICINN 2007-66665-C02-01, the INFN-MICINN
agreement program ACI2009-1038, consolider-ingenio 2010 program
CSD-2008-0037 and by CUR Generalitat de Catalunya grant 2009SGR502.
and by Conselho Nacional de Desenvolvimento Cient\'ifico e
Tecnol\'ogico (CNPq) and Funda\c{c}\~ao de Amparo \`a Pesquisa de
Estado de S\~ao Paulo (FAPESP).


\begin{thebibliography} {99}
\bibitem{anomalous} For a review see: H.\ Aihara {\it et al.}, {\it
    Anomalous gauge boson interactions} in Electroweak Symmetry
  Breaking and New Physics at the TeV Scale, edited by T.\ Barklow,
  S.\ Dawson, H.\ Haber and J.\ Seigrist, (World Scientific,

\bibitem{tevatron}
  V.~M.~Abazov {\it et al.}  [D0 Collaboration],
  arXiv:0907.4952 [hep-ex].
%
  T.~Aaltonen {\it et al.}  [CDF Collaboration],
  Phys.\ Rev.\ Lett.\  {\bf 104}, 201801 (2010)
  [arXiv:0912.4500 [hep-ex]].


\bibitem{lep}
  S.~Schael {\it et al.}  [ALEPH Collaboration],
  Phys.\ Lett.\  B {\bf 614}, 7 (2005);
%
  P.~Achard {\it et al.}  [L3 Collaboration],
  Phys.\ Lett.\  B {\bf 586}, 151 (2004)
  [arXiv:hep-ex/0402036];
%
  G.~Abbiendi {\it et al.}  [OPAL Collaboration],
  Eur.\ Phys.\ J.\  C {\bf 33}, 463 (2004)
  [arXiv:hep-ex/0308067].
%
  J.~Abdallah {\it et al.}  [DELPHI Collaboration],
  Eur.\ Phys.\ J.\  C {\bf 54}, 345 (2008)
  [arXiv:0801.1235 [hep-ex]].



\bibitem{Amsler:2008zzb}
  C.~Amsler {\it et al.}  [Particle Data Group],
  Phys.\ Lett.\  B {\bf 667}, 1 (2008).

\bibitem{indirect} See, for instance,
  S.~Alam, S.~Dawson and R.~Szalapski,
  Phys.\ Rev.\  D {\bf 57}, 1577 (1998)
  [arXiv:hep-ph/9706542];
  S.~Dawson and G.~Valencia,
  Phys.\ Lett.\  B {\bf 333}, 207 (1994)
  [arXiv:hep-ph/9406324];
  O.~J.~P.~Eboli, M.~C.~Gonzalez-Garcia and S.~F.~Novaes,
  Mod.\ Phys.\ Lett.\  A {\bf 15}, 1 (2000)
  [arXiv:hep-ph/9811388];
  O.~J.~P.~Eboli, S.~M.~Lietti, M.~C.~Gonzalez-Garcia and S.~F.~Novaes,
  Phys.\ Lett.\  B {\bf 339}, 119 (1994)
  [arXiv:hep-ph/9406316];
  S.~Dawson and G.~Valencia,
  Phys.\ Lett.\  B {\bf 333}, 207 (1994)
  [arXiv:hep-ph/9406324].

\bibitem{Dobbs:2005ev}
  M.~Dobbs,
  AIP Conf.\ Proc.\  {\bf 753}, 181 (2005)
  [arXiv:hep-ph/0506174].

\bibitem{Hagiwara:1986vm}
  K.~Hagiwara, R.~D.~Peccei, D.~Zeppenfeld and K.~Hikasa,
  Nucl.\ Phys.\  B {\bf 282}, 253 (1987).


\bibitem{madevent} 
  F.~Maltoni and T.~Stelzer,
  JHEP {\bf 0302}, 027 (2003)
  [arXiv:hep-ph/0208156].

\bibitem{cteq6l}
  J.~Pumplin, D.~R.~Stump, J.~Huston, H.~L.~Lai, P.~M.~Nadolsky and W.~K.~Tung,
  JHEP {\bf 0207}, 012 (2002)
  [arXiv:hep-ph/0201195].


\bibitem{linear}
W.\ Buchm\"uller and D.\ Wyler, 
  Nucl.\ Phys.\ B {\bf 268}, 621 (1986).
K.\ Hagiwara, S.\ Ishihara, R.\ Szalapski and D.\ 
  Zeppenfeld, Phys.\ Lett.\ {\bf B283} (1992) 353;
  Phys.\ Rev.\ {\bf D48} (1993) 2182.

\bibitem{Kum} 
  J.~Kumar, A.~Rajaraman and J.~D.~Wells,
  Phys.\ Rev.\  D {\bf 78}, 035014 (2008)
  [arXiv:0801.2891 [hep-ph]].


\bibitem{Han} 
  T.~Han and Y.~Li,
  Phys.\ Lett.\  B {\bf 683}, 278 (2010)
  [arXiv:0911.2933 [hep-ph]].


\bibitem{unitarity}
  U.~Baur and D.~Zeppenfeld,
  Phys.\ Lett.\  B {\bf 201}, 383 (1988).

\bibitem{unitWW}
  M.~Suzuki,
  Phys.\ Lett.\  B {\bf 153}, 289 (1985);
  C.~Bilchak, M.~Kuroda and D.~Schildknecht,
  Nucl.\ Phys.\  B {\bf 299}, 7 (1988);
  G.~J.~Gounaris, J.~Layssac, J.~E.~Paschalis and F.~M.~Renard,
  Z.\ Phys.\  C {\bf 66}, 619 (1995)
  [arXiv:hep-ph/9409260];
  G.~J.~Gounaris, F.~M.~Renard and G.~Tsirigoti,
  Phys.\ Lett.\  B {\bf 350}, 212 (1995)
  [arXiv:hep-ph/9502376].


\bibitem{qcdnlo}
  U.~Baur, T.~Han and J.~Ohnemus,
  Phys.\ Rev.\  D {\bf 51}, 3381 (1995)
  [arXiv:hep-ph/9410266];
  U.~Baur, T.~Han and J.~Ohnemus,
  Phys.\ Rev.\  D {\bf 53}, 1098 (1996)
  [arXiv:hep-ph/9507336].

\bibitem{ewnlo}
  E.~Accomando and A.~Kaiser,
  Phys.\ Rev.\  D {\bf 73}, 093006 (2006)
  [arXiv:hep-ph/0511088].
\bibitem{anonlo}
F.~Campanario, C.~Englert and M.~Spannowsky,
  Phys.\ Rev.\  {\bf D82 } (2010)  054015.
  [arXiv:1006.3090 [hep-ph]].

\end{thebibliography}
\end{document}